# Photostrictive materials

B. Kundys

*Institut de Physique et de Chimie des Matériaux de Strasbourg (IPCMS), UMR 7504 CNRS-ULP, 67034 Strasbourg, France*

Light-matter interactions that lead to nonthermal changes in size of the sample constitute a photostrictive effect in many compounds. The photostriction phenomenon was observed in four main groups of materials, ferroelectrics, polar, and non-polar semiconductors, as well as in organic-based materials that are reviewed here. The key mechanisms of photostriction and its dependence on several parameters and perturbations are assessed. The major literature of the photostriction is surveyed, and the review ends with a summary of the proposed technical applications.

## TABLE OF CONTENTS

## INTRODUCTION

Shape responsive solids that are sensitive to a variety of external excitations upon variation of environmental conditions constitute an important class of functional smart materials.[1–3] Their subclass, involving light-matter interactions that result in non-thermal sample deformation, is termed photostrictive. The photostriction differs in origin depending on the type of investigated material. The photostriction in electrically polar solids can be defined as photo-induced deformation of the lattice associated with a change in the internal electric field leading to a converse piezoresponse in the photovoltaic compounds. In organic polymers, light can trigger a transformation in molecular structure within the same chemical formula inducing large volumetric changes. In polar semiconductors, light modulates an electric field via free charge generation at the surface and causes an elastic strain due to the converse piezoelectric effect. In non-polar semiconductors, such as Si or Ge light irradiation, usually larger than a band gap creates an excess of electron-hole pairs in the conduction band, leading to deformation of the sample directly or via a change in atomic bonds in covalently bonded amorphous semiconducting materials.

This paper reviews the work on photostriction since its discovery in these groups of materials, with a concern to unite, compare, and summarize the existing disparate findings and analyses that describe the photostrictive phenomenon.

## FERROELECTRIC MATERIALS

The discovery of the photostriction phenomenon in electrically polar compounds dates back to the sixties, when the Japanese paper of Tatsuzaki *et al*. reported a photo-induced strain in single crystals of SbSI.[4] This work was then followed by other authors investigating further the same material.[5–7] Afterward, investigations were focused on the light-induced ferroelectric bimorph bending by Brody.[8,9] Most of the studies including $Pb[Zr_xTi_{1-x}]O_3$ (PZT-based) materials development and device designing for applications, belong, however to the group of Uchino. The renewed attention is currently devoted to $BiFeO_3$ (BFO) and $PbTiO_3$ ferroelectric compounds. The background mechanism of photostriction in ferroelectrics is a combination of a bulk photovoltaic effect,[10] i.e., substantial voltage generation and converse piezoelectricity.

## SbSI single crystals

The SbSI single crystals were pioneering objects for photostriction measurements among ferroelectric compounds.[4] This compound is at the same time photosensitive[11] and ferroelectric[12] with a Curie point of 22 °C. The wavelength of maximum photoconduction lies between 580 and 740 nm depending on the stoichiometric ratios between $Sb_2S_3$ and $SbI_3$. It was reported that the length of an SbSI crystal along the polar **c** axis is changed when it is illuminated uniformly by visible light, in the presence of a dc electric field along the polar c axis. The reported photostriction change (order of $10^{-5}$) was found to be opposite to that of warming excluding ordinary thermal expansion due light absorption. Moreover, the photoinduced strain was found to be positive below the $T_c$ and negative in some narrow temperature region just above $T_c$ . (Ref.4)   With further heating, no photostriction was observed in the paraelectric phase confirming its relation to the electrically polar state only. The maximum photostriction was observed near $T_c$ in a polar phase.

Notably, the observed photostriction shows a discontinuity as a function of temperature near $T_c$, passing through positive and negative maxima with a large electric field dependence

This effect was also found to be wavelength dependent by other authors[13] with amonotonous dependence shown on Fig. 3(a).

The unexpected wavelength dependence of photostriction correlates with the photoconductivity for both ferroelectric (15 °C) and near-paraelectric (25 °C) states (Fig. 3(b)). The sign change for both states in photostriction is observed. At the near-paraelectric (25 °C) state, the observed wavelength dependence was proposed to be predominantly a consequence of electric field screening by light generated charges as an electric polarization averages to zero in the near-paraelectric state. In the ferroelectric region, however, the wavelength dependence results from two competitive contributions: electric field redistribution leading to increment of electric field and electric field screening by the light generated charges leading to decrease of electric field in the sample. This suggestion seems to be in agreement with the Landau free energy expansion analysis[14] with two competing contribution: the fist contribution is proportional to square of polarization, while second one is proportional to the average light-induced charge concentration (see the model description). Obviously, the two contributions opposing each other as a function of wavelength give rise to the observed amonotonous dependence. In confirmation of this assumption, the photostriction in the ferroelectric state changes its sign at ∼650 nm where maximum of photocurrent is observed (Fig. 3).

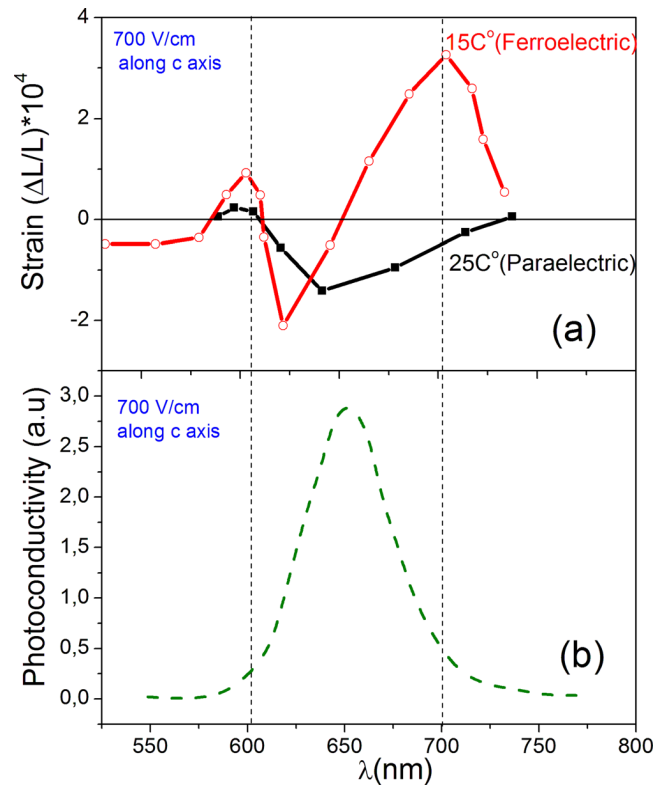

FIG. 3. (a) Photostriction due to illumination in ferro and near para-electric (just after the transition) states. Silver electrodes were used. (b) Photoconductivity as a function of wavelength after Ref. 13.

Despite its extraordinary photovoltaic and photostrictive properties, the SbSI single crystals are not attractive for room temperature applications due to low ferroelectric critical temperature ($T_c \approx 295$ K). In that regard, the possibility of $T_c$ increase towards 330 K in the modified SbSI ceramics[15–18] can stimulate a farther research in this direction.

### Modified by lanthanum Pb(Zr$_x$Ti$_{1-x}$)O$_3$ ferroelectric ceramics

The photostriction studies were further focused on the room temperature ferroelectric lanthanum modified PZT (PLZT) by Brody[8] and Uchino.[19–22,27] As expected for ferroelectric ceramics, the preparation methods used to synthesize samples affect largely the photostrictive properties due to changes in homogeneity, stoichiometry, and grain sizes.[23] Remarkably, with grain size optimization, the photostriction as high as 0.01% was reported to be achievable.[24] Such a large value is very encouraging and already comparable to electrical field-induced strain of common piezoelectric materials. Due to the light penetration depth dependence, the photostriction logically increases as sample thickness of the WO$_3$ doped PLZT ceramics decreases[25] with an optimal sample thickness determined to be 33 $\mu$m. Moreover, the increase of photoelastic effect can be expected for coherent light using the model of validation and analysis.[26] Unfortunately, the response time being tens of seconds[23] is still slow due to high dielectric permittivity values in PZT based compounds. This issue can be understood via simple empirical formula for linear photostriction as a function of time ($t$), piezoelectric coefficient $d_j$, and light-induced electric field $E_j$[28]

$$\lambda_{hv} = \frac{\Delta L}{L} = d_j E_j (1 - e^{-t/RC}), \quad (1)$$

where R and C are resistance and capacitance of the sample. For $t \ll 1$, one can write $\Delta L/L \cong d_j E_j(t/RC)$. Where low dielectric permittivity (capacitance) is preferable to have faster response speed. By contrast, large piezoelectric coefficient and large internal electric field (polarization) are needed for large photostriction. The PLZT exhibits a large photostriction under near-ultraviolet (UV) illumination due to its wide band-gap. Although no direct wavelength dependent photostriction measurement is present in the literature, the maximum photostriction can be expected at the 366 nm wavelength where the maximum photovoltaic properties in the PLZT samples[29] are observed.

### The Sn$_2$P$_2$S$_6$ crystals and ceramics

The sulfide ferroelectric Sn$_2$P$_2$S$_6$ with a band gap near 2.3 eV undergoes a second order[30] ferroelectric phase transition at 339 K with the crystal symmetry change from polar monoclinic P$_c$ to nonpolar P$_{21/c}$.[31,32] A prominent photoconductivity,[33] large piezoelectric coefficients, and electro-optic effects[34] made this compound a candidate for photostriction studies.[35] The photovoltaic currents were found to be 5 times smaller for ceramics than for single crystals, directly correlating with the value of the electric polarization. Using the general formula (1) where photostriction is a product of light generated electric field and piezoelectric coefficient, the photostriction was evaluated from the current voltage characteristic under infrared light illumination. The photostriction estimate for single crystals gives a quite small value of 0.0047 ppm under the infrared light intensity of 2.1 W/cm$^2$.

### The BiFeO$_3$ single crystals and films

Multiferroic BiFeO$_3$ compound is an extraordinary system containing several built-in order parameters that can lead to exotic cross-functionality. In addition to electromagnetic effect,[36] it shows also optical functionalities[37] that, thanks to ferroelectricity, can interact with strain.[38] As mentioned above, the photostriction in ferroelectric materials is known to originate indirectly from the superposition of photovoltaic and strictive effects resulting in a very slow response time (typically tens of seconds) in classical ferroelectrics.[23] In this respect, the report on the fast (below 100 $\mu$s) photostriction in a single crystal of BiFeO$_3$ has recently attracted an attention from the optical scientific community.[39–43] Because photostriction depends on the light penetration depth, its response time and a magnitude can still be improved in thin films. Indeed, the photostriction response in BFO films was reported to be in the picoseconds range.[39] As the electric polarization and piezoelectric coefficients play an important role to determine the magnitude of the effect (Eq. (1)), the fast photostriction in BFO films can yet be improved via doping induced polarization increase[39] and enhancement of the piezoelectric coefficient. For example, BFO doping with Tb,[44] Sm,[45] or La,[46] was reported to increase piezoelectric coefficients significantly. Taking into account recently reported quasi-instantaneous photostriction response,[47] the exciting opportunity to design fast responsive light-controlled devices can be realized. As light energy decreases, the photostriction effect in a single crystal of BFO also shows amonotonous dependence[48] and follows the photocurrent reported in Ref. 49 with a some delay (Fig. 4). This feature appears to be common for polar semiconductors also (see the Polar Semiconductors section) and is connected with amonotonous light-induced charge excitation with an explanation similar to that of SbSI.

Multiferroic BFO also possesses a magnetic degree of freedom as it is antiferromagnetic. This can potentially couple photostriction to magnetic order via elastic interactions. Interestingly, the photostriction in BiFeO$_3$ single crystal was found to decrease in magnetic field.[38] Although this effect was checked not to be connected with magnetostriction or visible magnetic field induced changes in polarization, its exact origin remains to be investigated. As discussed above the photostriction may consist of two competing mechanisms. Indeed, the evidence for both optical rectification[39] and light-induced change generation[40] mechanisms of photostriction in BiFeO$_3$ were reported. However, the free charge contribution is under debate.[47] As far as thin films are concerned, it is also important to sort out any possible stray free charges as leakage currents in thin films of BFO, screening the electric polarization, are known to be a serious challenge.[50,51]

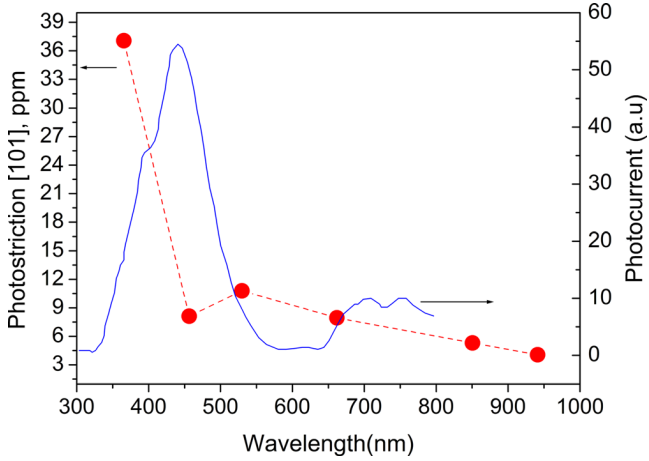

FIG. 4. Wavelength dependence of photostriction response in the single crystal of BiFeO$_3$ (Ref. 48) and photocurrent of BiFeO$_3$.

### PbTiO$_3$ thin films

The fast (tens on ps time scale) photostrictive effect has also been recently seen in thin films of the prototypical ferroelectric PbTiO$_3$ and explained also to result from the coupling to its intrinsic photovoltaic response.[52] The light-induced changes in the unit-cell tetragonality were observed using the time-resolved X-ray scattering reaching 0.25% of maximum of photoinduced strain. The photostriction depends nonlinearly on the light fluence (Ref.52) showing a saturation that is inconsistent with dominant thermal effects.

After ultrafast pulse excitation of 50 fs at 400 nm, the changes in tetragonality lasting several nanoseconds are observable. The concluding mechanism was proposed to originate from the coupling of the photogenerated changes in the internal electric fields to the converse piezostrain within the film. This is consisted with the qualitative theory described above where the wavelength dependence of photostriction and its correlation with photocurrent is a key in determining dominant contribution to the photoelastic effect.

### A model description of photostriction in ferroelectric materials

While formula (1) is successful in describing qualitatively the photostriction and its response time dynamics at saturated ferroelectric state, it also assumes that the piezoelectric coefficient is light independent that strictly speaking not always the case. An analytical description containing the strain-induced effects involves additional terms in describing the light-induced charge generation $\sum_j n_j \varepsilon_j \Omega_j$ to the free energy of the system[14]

$$F = \frac{1}{2}\alpha P^2 + \frac{1}{2}\beta P^4 + \frac{1}{2}\gamma P^6 - \frac{1}{2}\sum_i \sum_j k_{ij} \Omega_i \Omega_j - P^2 \sum_j \xi_j \Omega_j + \sum_j n_j \varepsilon_j \Omega_j, \quad (2)$$

where $k_{ij} = \partial^2 F/\partial\Omega_i\partial\Omega_j$ components of the elastic stiffness tensor. $\Omega_{ij}$ is the strain tensor. $\xi_j = \partial^3 F/\partial P^2 \partial\Omega_j$ is the electrostriction tensor. $\alpha$, $\beta$, and $\gamma$ are the known coefficients of ferroelectric free energy expansion. $n_j$ is an average concentration of electron subsystem charges at the energies levels $\varepsilon_j$ (conduction band levels of traps and recombination) close to the bandgap energy. Taking the derivative with respect to polarization, which is the internal electric field, one can get

$$\frac{\partial F}{\partial P} = \alpha P + \beta P^3 + \gamma P^5 - 2P \sum_j \xi_j \Omega_j = E_j. \quad (3)$$

The components of the deformation tensor $\tau_j$ can be defined as $\partial F/\partial\Omega_j$

$$\frac{\partial F}{\partial \Omega_j} = -\frac{1}{2}\sum_i k_{ij}\Omega_i - P^2 \xi_j + \sum_j n_j \varepsilon_j = \tau_j. \quad (4)$$

Putting $\Omega_l = 0$ in Eq. (1), one can simplify the equation to

$$\tau_j = \sum_j n_j \varepsilon_j - P^2 \xi_j. \quad (5)$$

This equation shows that the deformation does not depend on the sign of polarization as expected and should change in the presence of the generated charges. On the other hand, the piezoelectric coefficient determined from Eq. (1) using Eqs. (3) and (4) becomes

$$d_j = \frac{-P^2 \xi_j + \sum_j n_j \varepsilon_j}{\alpha P + \beta P^3 + \gamma P^5 - 2P \sum_j \xi_j \Omega_j}. \quad (6)$$

Importantly, this free energy expansion analysis shows that the photostriction (Eq. (5)) itself has two competing contribution: one can arise from direct change in polarization (for example, via optical rectification effect) and another from the light generated charges. Such a competition can indeed be confirmed by experimental results on the wavelength dependence obtained for SbSI single crystals (Fig. 3(a)) and therefore manifests a qualitative agreement. However, a quantitative verification of the theory performed on SbSI single crystals has not been found to agree with experimental results.[14] It was suggested that changes in ferroelectric domain state under illumination (electric field changes and leads to domain movement) must be taken into account. Indeed, in ferroelectric samples, the both piezo (odd with electric field) and electrostrictive (even with electric field) effects are present. Such an interplay together with charged domains movement at low subcoercive electric fields can lead to the exiting multilevel hysteresis behavior in both bulk[53] and surface[54] effects and the theory must be extended to include them.

## POLAR SEMICONDUCTORS

### CdS single crystals

First photostriction study in polar semiconductors started with the photomechanical effect observation in noncentrosymmetric CdS single crystals by Lagowski and Gatos.[55] It was observed that a visible illumination induces an elastic deformation resulting in the mechanical vibrations of CdS cantilevers with an orientation perpendicular to the crystallographic axis (001) with the two types of possible deformations (Fig. 6). Since the cantilevers are normally longer than wider, one can assume that the deflection of the type shown on Fig. 6 (left side) is dominant. Under such assumption, one can perform an estimation of the magnitude of the photostriction effect. For small deflection ($\Delta$), large radius ($R$) of deformation and assuming that $\alpha \approx L/R$, where $L$ is the initial sample length one can write

$$\Delta \cong L^2/2R. \quad (7)$$

Taking the reported[55] relation of deflection of the free end of the cantilever to its length (10 mm) of $5 \times 10^{-2}$, the radius of the curvature estimate gives approximately 100 mm. The photostriction can then be calculated using the cantilever thickness (t) and the bending radius via the relation

$$t/2R \quad (8)$$

and gives approximately 75 ppm.

The effect was explained to originate from the combination of light-induced changes to the surface electric field and elastic strain due to the converse piezoelectric effect. Although the wavelength dependence of the amplitude of vibrations exhibited by the CdS wafers was observed very similar to their surface-photovoltage spectrum, it was found different considerably from the photoconductivity wavelength dependence (Fig. 7). This issue is in agreement with experimental observations in SbSI single crystals (Fig. 3) originating from the possible interplay of the two competing mechanisms in the photoelastic effect (Eq. (5)).

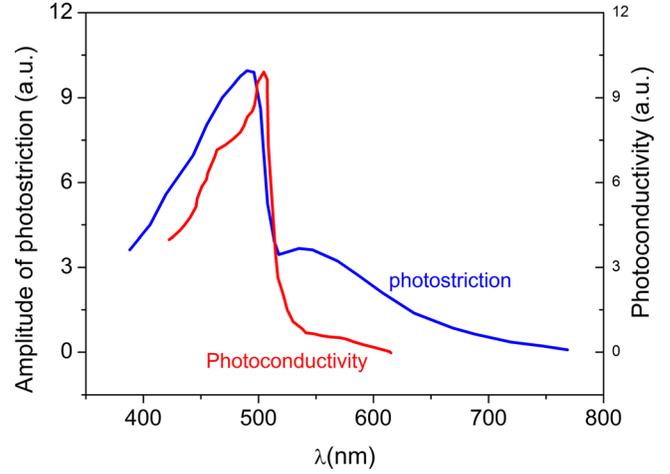

FIG. 7. Wavelength dependence of photostriction for CdS cantilever and its correlation with the photoconductivity spectrum.

### GaAs single crystals

Due to its noncentralsymmetric crystallographic structure, the similar photomechanical vibrations were also reported for (111) GaAs single crystals.[56] With the chopped light, the cantilever was excited to their resonant vibration. As in the case of CdS photostriction effect, the wavelength dependence of the amplitude of this vibration was found to be different to that of photoconductivity (Fig. 8). The results were interpreted within a model based on light-induced modulation of the piezoelectric surface stresses. The amplitude of the vibration was found to increase with decreasing pressure, apparently due to a decrease of the damping factor.

For incident photon energies larger than the bandgap, the amplitude of the photomechanical vibration versus illumination intensity exhibited a saturation that can be considered typical for surface photovoltage behaviour (Fig. 8 (inset)). The similar slope showing photostriction saturation versus light intensity was recently observed for $PbTiO_3$ ferroelectric film (see above) and is inconsistent with dominant thermal heating effect. Importantly, no photostriction was observed in the both CdS and GaAs wafers with prismatic surfaces. For both (001) CdS and (111) GaAs wafers, it was concluded that contributions from thermoelastic or pyroelectric effects can rather be excluded due to low intensity of sub-bandgap monochromatic radiation. In addition, the amplitude of the photomechanical vibration was found to decrease significantly under superimposed steady-state illumination. Interference by light pressure effects was also excluded since the photostriction found only when surfaces perpendicular to the polar axis are illuminated. It was therefore concluded that photostriction should occur in other non-centrosymmetric semiconductors with a depletion surface layer. Furthermore, it can also exist in the depleted regions in the bulk materials and, for example, in rectifying junctions perpendicular to the polar axis. The rough estimation of the photostriction in the GaAs wafers for 10 μm thick sample employing analysis shown on Fig. 6 and data from Ref. 56 gives 0.4 ppm, which is more than 2 orders of magnitude less than for CdS wafers. The possible explanation of this difference is the large difference in the absolute value of the piezoelectric coefficients of these two compounds[57,58] in agreement with empirical formula (1).

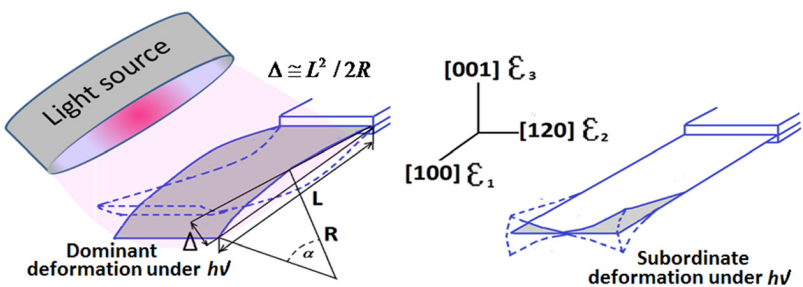

FIG. 6. Crystallographic orientation of the CdS cantilever with two possible types of light-induced deformations.

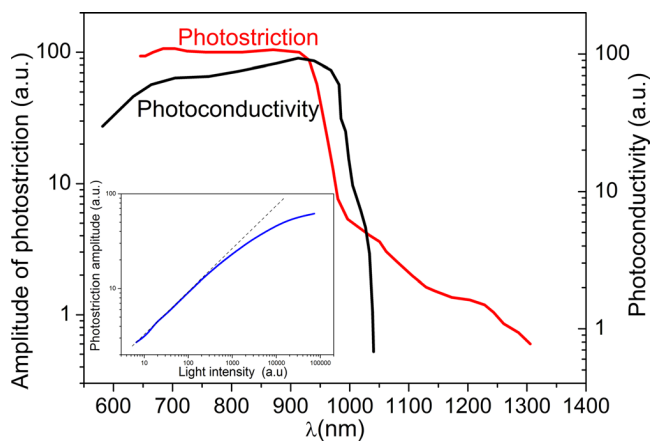

FIG. 8. Wavelength dependence of photostriction for (111)GaAs cantilever and its correlation with the photoconductivity spectrum. Inset shows photostriction amplitude as a function of the light intensity for the light 1.45 eV.

## NONPOLAR SEMICONDUCTORS

### Germanium

Generated by monochromatic light excess electron hole pairs were reported to create the lattice dilatation in nonpolar semiconducting germanium.[59] Enlightened semiconductor at the wavelength corresponding to photon energies larger than the band gap develops electron-hole pairs throughout the surface at the thickness of the light penetration depth. The light generated electrons in the conduction band can contribute to the atomic bonding energy, whereas holes in the valence band can decrease the energy of covalent bonds. Because a volume of semiconductor can depend on the occupation of the electronic energy levels,[60] the material can deform in the illuminated region. The sign of the photostriction is determined by the pressure susceptibility of the energy gap which is positive in germanium leading to lattice dilation.[61–63] The spectral response of the photostriction and photoconductivity was measured in Ref.59 As energy decreases, the photostriction follows the photoconductivity with a small delay. Notably, this behavior is inverse to the one observed for polar semiconductors and ferroelectrics (see Figs. 4, 7, and 8) indicating a different photostrictive origin. To explain this feature, the existence of the $\langle 100 \rangle$ exciton states with relatively long apparent lifetimes has been postulated.[59]

The photostriction was found to be a linear function of the change carriers concentration with the maximal reported value of $1.12 \times 10^{14}$ cm$^{-3}$. Taking this value, the photostriction can be estimated to be $7.84 \times 10^{-4}$ ppm for the reported sample thickness of 0.5 mm that is three orders of magnitude smaller than in the low photostrictive polar GaAs semiconductor.

### Silicon

Contrary to the light-induced lattice dilation in germanium, a light-induced contraction effect was observed in phosphorus-doped n-type silicon ((111) face).[64] The response of the strain quartz sensor glued on the Si crystal due to laser light of 1.17 eV with a pulse duration of 25 ns has shown the behavior opposite to that expected for thermal expansion implying the initial lattice contraction. A similar explanation of this phenomenon in terms of a light-induced charge generation was proposed. The contraction agrees well with the negative sign of pressure susceptibility of the energy gap, which is indeed negative for silicon. Photoexcitation of electron-hole pairs therefore contracts the lattice under the illumination in silicon. Similar as for germanium, the photostriction was found to be linear function of light intensity. The magnitude of photostriction at room temperature can be evaluated to be $-0.414$ ppm for the light fluence of 0.03 J/cm$^2$ using data from Ref. 64. The photostriction studies in the Si cantilevers have also revealed a linear dependence upon light power (not shown) and wavelength (Ref. 65). The photostrictive deflection increases linearly with increasing wavelength up to the drop at ~1.1 μm.

Using data from Ref. 65 and Eqs. (7) and (8), one can estimate the photostriction of ~1.25 ppm at the 780 nm excitation for the reported sample thickness of 0.5 μm. Yet, six times larger effect was found by time resolved X-ray diffraction measurements at 123 K using much larger light intensity.[66] Observing the shift of a Bragg peak during laser illumination at 123 K, the photostriction of 6.4 ppm has been found for the sample thickness of 0.5 mm with the fluence of 127 mJ cm$^{-2}$ at the wavelength of 2480 Å.

### Carbon nanotubes

Semiconducting carbon nanotubes possess a wide range of direct bandgaps in the solar spectrum range, and high carrier mobility at low scattering ratio, which make themselves ultimate photovoltaic materials.[67] Especially, single wall carbon nanotubes (SWNT) manifest a record power conversion efficiencies in the form of p-n junctions with Si.[68] The first observation of photostrictive effect in low dimensional carbon structures was reported by Zhang and Iijima[69] and was found to be too large to be explained by thermal expansion alone. It was concluded that electrostatic interactions are important in the observed phenomenon. For low dimensional samples, the photon pressure can be important but was sorted out by the fact that the photostrictive-like movement direction was different from the light propagation direction in many cases. The white light with an intensity of 20 mW/cm$^2$ induced a displacement of 170 μm of the 1 mm long filament with 10 μm diameter in a time scale of 100 ms. However, the exact evaluation of the photostrictive contribution requires comparison with thermal expansion measurements. To address this issue, the SWNT suspended in air were studied by other authors[70] who determined that electron-phonon coupling explains the experimental photostrictive observations, neglecting thermal expansion. In contrast, a dominating thermal expansion was concluded for nanotubes in surfactant or in bundles. The photostriction of the both SWNT and multi wall carbon nanotubes (MWNT) was studied in the form of free standing thin films in Ref. 71. It was concluded that the photostriction is orders of magnitude larger than the expected thermal expansion effect. For MWNT films, the 675 nm light at the irradiance of 1.5 mW/cm$^2$ provokes an initial subordinate contractive deformation with the stress change of −5.5 kPa in 1 s and then expansion is observed with the maximum of photoinduced stress of ~0.3 MPa for longer time scales. Assuming the smallest estimated Young's modulus value of 270 GPa (Ref. 72) and zero thermal expansion contribution, the photostrictive deformation is ~1.1 ppm. The SWNT films, however, have shown a dominant light-induced contraction without deformation sign change with a smaller effect (−57.3 kPa). More generally, the carbon nanotubes in different composites structures constitute a promising class of photoactuation systems that are in detail reviewed in Ref. 73.

### Chalcogenide glasses

The covalently bonded amorphous semiconducting materials containing considerable amount of chalcogen atoms were reported to show photostriction in 1974.[74] Contrary to abovementioned elastic photostriction in classical semiconductors here light-induced deformation has a plastic character and persist after light is off. Only a thermal heating above certain critical temperature can recover the initial sample shape. The phenomenon was first shown in thin films of $As_2Se_3$, $As_2S_3$, $As_{40}Se_{50}Ge_{10}$, $As_{40}Se_{25}S_{25}Ge_{10}$, and $As_{40}Se_{10}S_{40}Ge_{10}$ deposited in the mica substrates that were found to bend under heating and become flat under light. The deformation level was evaluated from the change in the radius of bending to be of order of $10^{-3}$. Unfortunately, the process of the light-induced deformation is slow (2 h timescale). While As-based samples show positive photostriction, the Ge based glasses ($GeSe_2$ and $GeS_2$) show light-induced contraction.[75] The photostriction reported $As_2S_3$ films range from 0.42% (Ref. 76) to ~6% (Refs. 77 and 75) in response to light illumination of a bandgap energy. Yet, a possibility of an order of magnitude improvement in photostriction has been reported for $As_xS_{100-x}$ ($5 \leq x_{As} \leq 40$) bulk glasses[78] and $As_2S_3$ thin films.[79] The photostriction in these materials was reported to be nonthermal in origin[79,80] as well as the optomechanical response of the thin film cantilever (configuration similar to the one shown in Fig. 6) was reported to be accurately controlled by varying the angle between light polarization and the cantilever axis.[81]

Several mechanisms were proposed[82,83,88] to describe photostriction in the chalcogenide glasses and it seems to be reasonable to assume that the phenomenon is generally determined by light efficiency in modifying the bond between atoms in such a way that it finds another equilibrium state leading to a different atomic configuration. This scenario is possible thanks to amorphous nature of the chalcogenide glasses and is usually accompanied with a optical absorption edge shift (photodarkening effect).[84–87] Although the exact mechanism of the light-induced change in the atomic bond configuration is still unclear to a large extent, it seems that light generated charges exited from the valence to conductance band are responsible for atomic bond changes.[88] Indeed, some experimental results indicate that the number of light generated charges is related to the photoinduced volume expansion.[80]

## ORGANIC POLYMERS

Unlike other photostrictive compounds, the organic based materials showing photostrictive effect are multiple. A photosensitive organic molecules resulting in the light-induced reorientation or ionization reaction are generally responsible for the photostriction mechanism in these materials. Because many organic based materials were already reviewed elsewhere,[89–91] we will only briefly list the most important compounds in which photostriction was reported to differ from direct light heating effect as many reports risk to be obscured by dominant thermal expansion.[92,93]

The first report on the photostriction effect in polymer dates back to 1971 when Van der Veen and Prins have reported photomechanical energy conversion in a polymer membrane.[94] These poly-(4,4′-diaminoazobenzenepyromellitimide) films were known to undergo isomerization transition under UV light and a volume photostrictive effect of 3.6% was observed (1.2% of linear contraction). The response and recovery times were in the range of tens of minutes that is correlated with the time needed for chemical structure transition. Although a heating contribution may be present, it was not found to be dominant as the independent measurements as a function of temperature had shown that contraction induced by light is much larger than a thermal deformation.

The related compound of poly(ethylacrylate) networks with azo-aromatic crosslinks shows an UV induced a geometrical change in the chemical structure of the containing azobenzene unit. These compounds in a form of film have shown photoelongation of 0.25% in 1980 (Ref. 95) due to so called light-induced photoisomerization transition.

The presence of azobenzene units can also trigger photostriction in the azobenzene-containing liquid-crystalline polymers (LCPs) via similar photochemical structure changes of $x^2$ containing molecules. While the molecular formula remains the same, the exact arrangements of the exited molecules change. The recent research progress in the area of photosensitive and photomobile applications is in detail reviewed in Ref. 90.

The polymer materials manifesting both viscosity and elasticity (elastomers) were reported to develop the largest photostriction—between 10% and 400%.[96] Molecular shape in these materials is tied to macroscopic sample shape and depends significantly on the state of nematic order. The background mechanism is a light-induced movement of a photosensitive cluster of atoms that changes a chemical bond and produces a twist in a polymer chain. Although the change in chemical bond is not directly connected with volume photostriction, it induces nematic order-disorder transition leading to the enormous photostrictive effect. An elegant demonstration of photostriction was also reported for crystalline elastomer films based on azobenzene derivative.[97] The fascinating bending of film repeatedly and precisely along any chosen direction was reported. Using the linearly polarized light, the bending can be controlled by light polarization rotation. The above mentioned photostriction effects are unfortunately in the seconds or minutes response time scales. In that respect, the recent reports on the ms[98] and even $\mu$s[99] photostrictive speed time scales should trigger a new research effort.

## SPIN-CROSSOVER MOLECULAR CRYSTALS

These molecular magnets[100] exhibit a high spin (HS, S = 2) to low spin (LS, S = 0) state transition that can be switched by pressure, temperature, or light irradiation.[101] Owing to their organic content and magnetic metal atoms, these materials can change their structural arrangements under light leading to the change in the magnetic state. The generally accepted mechanism of photostriction here is a light sensitive electronic subsystem that intrinsically lead to the change in the bond configuration to form another metastable structure with different elastic,[102] optical, and magnetic properties.[103]

In the representative iron (II) molecular complexes [Fe(phen)$_2$(NCS)$_2$], for example, the photostriction happens together with a change in magnetic properties and can reach volumetric change of 1.1% at 30 K (Ref. 104) in connection with the light-induced excited spin state trapping effect.[105,106] A spin state photoswitching was also reported to be followed by the volume expansion in the molecular Fe(III) spin-crossover monoclinic polymorph [(TPA)Fe(III)(TTC)]PF$_6$ with a different time scales for thermally and light-induced regimes.[107] Other photostrictive candidates of this rich family of compounds can be found in Refs. 101 and 103. Unfortunately, a temperature range where photostriction can be expected is well below the room temperature in most of these materials.

## APPLICATIONS

For the reason that photostrictive materials involve a coupling between optical and mechanical functionalities,

they can serve as convertors between light and mechanical energy. Hypothetically, photostrictive materials can be used ubiquitously where piezoelectric or magnetostrictive materials are used with an important wireless advantage and without a need of magnetic field. Although the relative photostrictive displacements at fast speed are small, their repetitive operation can, in principle, be used to develop acoustic or even electrical power. Up to this date, the most of the technical developments have been proposed for PZT based ceramics[26] and polymers based materials.[90] Here, we point out some motivating applications combining our suggestions with already reported ones, where the aim is to underline the potential advantages that photostrictive materials possess. The utilization of the photostrictive effect in already demonstrated and prospective technical applications includes as follows.

### Microactuation

Using a photostrictive bimorph configuration, a photodriven relay,[21] light-driven micro walking mechanism,[108] and photoacoustic devices[109] have been demonstrated on PZT based ceramics.[110] Other applications may be optically controlled electric delay lines, robots,[111] optical micro positioning, wireless control of flexible structures,[112] vibrations,[113] and photophones. In Ref. 20, it was also proposed to construct a sun chasing mechanisms for photovoltaic panels using photostrictive films on the flexible substrates. As far as polymers are concerned, light controlled biomimetic actuators become an interesting possibility as well as light assisted crack self healing effects in these materials.[114]

### Microsensing

Microsensing can include tunable sensors for incident radiation using strong photostriction dependence on the wavelength and power. So called indirect microsensors utilizing photodetection when the device is loaded with condensing moisture (microhygrometer) or incident interstellar dust (microbalance).[20] Since photostriction can depend on the electric field applied to the sample (Fig. 2) and polarization it can be also used to determine the polarization state without measuring charge or capacitance directly. The chemical gas sensing was proposed using a photostrictive cantilever device[115] utilizing the general cantilever approach for chemical and biological sensors.[116]

### Energy harvesting structures

The combination of photostriction materials with piezoelectric materials in multilayers or other hybrid structures may lead to light energy conversion into electricity similar to that reported for piezoelectric[117] and piezostrictive-magnetostrictive structures.[118] Adding magnetostrictive materials to such a structures can even offer an additional degree of freedom to control their performance by magnetic fields.

### Photonics

In optics, a fast and large photostriction can be a missing property to achieve all-optical control in optical logic elements.[119] For example, one can imagine photostrictive transistors that can control a light flow in a similar fashion as a transistor controls an electric current flow. There can be also a great deal of interest in substituting electro-optical components with photostrictive devices. For example, electrically controlled piezoelectric fibers[120] can be substituted by the polymer fibers[121] with a possible photostrictive effect for all optical operation. The performance is based on the possible optical control of the refractive index and birefringence because photostriction is closely linked to photoinduced refractive index change. For example, azobenzene containing polymers[90] and chalcogenide glasses Ge–As–Se,[122] $As_{60}Se_{40}$, and $Sb_{45}Se_{55}$[123] do show this property. As well as in $As_2S_3$ films, the photostriction is actually accompanied with an increase of refractive index.[124] Yet, many electrically polar and photovoltaic materials[125] and spin-crossover molecular crystals[126] do also show photorefractive effect. Consequently, holographic recording is another possible application.

### Strain mediated magnetization control

Controlling the magnetization direction in low dimensional magnetic materials is a key issue for magnetic recording. This is conventionally performed by applying external magnetic fields. In spintronics, for example, the magnetic field can be generated by passing large currents to micro strips generally made of gold. However, these currents generate heating effect that becomes increasingly important as dimensions are reduced below a micrometer, making this approach challenging for high-density magnetic memories. Use of photostriction here in a combination with magnetic overlayer having inverse magnetostriction effect can constitute an alternative and wireless approach. This idea is analogical to the piezoelectric magnetic anisotropy control via inverse magnetoelastic effect. The magnetization control was proven to be feasible using piezoelectric materials and recently it was also reported for photostrictive matrix with magnetoelastic elements.[127] The preliminary experiments with CoFe/BiFeO structures have also demonstrated a possibility of photoelastic-photoresistive-magnetoresistive coupling.[128] If a large and fast coupling is achieved it makes possible the integration of the light controlled magnetic properties with spintronics. The possible prototype memory tunneling magnetoresistive (TMR) cell is shown in Fig. 12. A highly magnetoelastic layer deposited onto photostrictive substrate changes its magnetic anisotropy in response to photostrictive effect of the substrate thanks to inverse magnetostrictive effect.[129] On the top of the magnetoelastic film, the tunneling layer is deposited to form TMR junction. Contrary to the lower ferromagnetic layer, the upper one must be weakly magnetostrictive to keep its magnetization insensitive to the light generated strain through the substrate. By the virtue of the TMR effect, this change can reflect itself in the vertical resistance change (Fig. 12(b)).

Taking into account that the bulk strain can show a well defined hysteresis loop behavior at subcoercive electric field region,[53] the multistate nonvolatile memory operation becomes possible. This should lead to an additional degree of freedom in hybrid straintronic-spintronic devices where stable changes in magnetization can be created by light and then maintained at zero power.

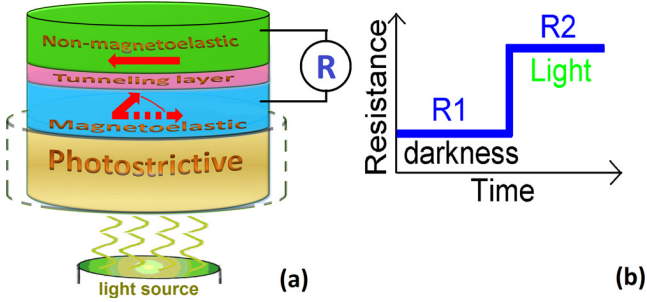

FIG. 12. TMR junction (a) and its expected light controlled operation (b). Red arrows show possible magnetization rotation as a function of elastic distortion induced by light.

**Photostrictive-magnetostrictive magnetometers**

The construction of photoelastic-piezomagnetic magnetometers can be another possibility. Its design involves a deposition of the magnetoelastic material on a top of the photoelastic plate (Fig. 13). A light beam with an alternating power or/and wavelength illuminated on the photoelastic plate can generate a photostrictive strain. This dynamic strain is transferred to the magnetoelastic layer. By the virtue of the inverse magnetostrictive effect, this dynamic strain generates a small magnetization rotation in the magnetic layer which can induce electromotive force in a pick-up coil. The low frequency signal field can then be extracted from a magnitude of the modulated electromotive force using conventional phase sensitive detection techniques.

Similar configurations involving a piezoelectric element have been used for low frequency magnetic field detection in Ref. 130 with a reported sensitivity of $6.9 \times 10^{-6}$ A·m$^{-1}$/$\sqrt{Hz}$. This value may now be improved owing to the technical progress in numerical processing in the phase sensitive detection. Moreover, it might be advantageous to modulate light power and frequency compared to an amplitude and frequency of the electric field in piezoelectric samples, often limited by electric breakdown and fatigue.

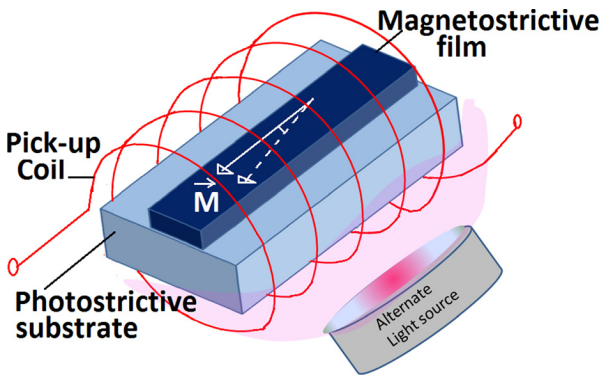

FIG. 13. Schematic diagram of the magnetostrictive-photostrictive sensor for a DC magnetic field detection. White arrows show possible magnetization modulation as a function of elastic distortion induced by light.

**Sonic–ultrasonic emission**

Sonic–ultrasonic emission induced by light is another possible application. These devices take advantage of their linear response under oscillation of light frequency or/and power. In this case, magnetostrictive/photoelastic multilayers may respond to both magnetic field and light, possibly leading to enhanced low frequency sonic emission. The currently used piezoelectric materials are limited by breakdown voltage limit, providing a room for magnetostrictive/photoelastic components improvements. Recently reported ultrafast GHz coherent shear phonons photogeneration and photodetection in the BiFeO$_3$ with an ultrahigh efficiency may contribute to an essential breakthrough in this direction.[131]

**Light controlled gas storage**

The use of photostriction effect in porous organometallic materials for gas storage is particularly interesting, owing to the high technological importance of this topic and the multi-fold energy (hydrogen) storage impact. Due to the peculiar structure of this class of materials, the volume appears to be a key parameter for the storage properties, controlling the pore size. The hydrogen binding energy is extremely sensitive to the inter-atomic distances, due to the steep dependence of the van der Waals interaction on atomic separation, thus providing a possible way for deformation dependent control. For example, strain effects have been found to affect dramatically the gas storage properties in Mg.[132] The key challenge here is to elaborate the porous materials with large photostriction at room temperature opening a way to fast optical pore size control.

**Concluding remarks and outlook**

We reviewed the current research progress in the field of photostriction and photostrictive materials touching some basics of the photoelastic phenomenon, related compounds, and mechanisms for light triggered strain control, with particular emphasis on finding the general rules in each exacting group of materials and comparing the magnitude of the effect. Table I summarizes the reported maximal values of photostriction for different compounds. For the reason that photostriction differs intrinsically from compound to compound and depends on thickness and light intensity, one can introduce a photostrictive efficiency at the given wavelength as

$$\eta_{eff} = t * \frac{\lambda_{h\nu}}{I}, \qquad (9)$$

where $t$ is the thickness along the illumination direction, $\lambda_{h\nu} = \Delta L/L$ is a linear photostriction, and $I$ is a light irradiance. Due to the linear approximation, the values of photostrictive efficiency have a rather indicative character, and the possible surface roughness dependence[133] is not taken into account by Eq. (9). The available data nevertheless show that the optimized PLZT ceramics tend to have the best photostrictive efficiency among the reported photostrictive ferroelectrics. The second place, however, belongs to the BiFeO$_3$ single crystal which might be more suitable for applications due to the much faster response time. The BiFeO$_3$ thin films are more efficient than the PbTiO$_3$ ones probably due to the unsaturated photostriction slope versus light fluence in the case of the BFO film.[47] In both groups of materials, ferroelectrics and polar semiconductors, the photostrictive mechanism is basically the same, involving direct light coupling to polarization and a possibility of light-induced free charge generation. Consequently, the exact contribution of each mechanism can depend on the excitation wavelength and can qualitatively be described by the Landau formalism with light generated charges term. In these materials, a general feature is the change in the photostriction trend versus wavelength that is correlated with a maximum in the photoconductivity.

TABLE I. Comparison of photostrictive properties for different compounds.

| Compounds | | Excitation (nm) | Thickness along illuminated direction | Light irradiance/ fluence | Maximum $\lambda_{h\nu}$ reported (%) | $\eta_{eff}$ ($m^3$/W) | References |
|---|---|---|---|---|---|---|---|
| Ferroelectrics | SbSI single crystals | 450 | … | … | −0.007 | … | 4 |
| | PLZT ceramics | 365 | 0.5 mm | 150 W/$m^2$ | 0.01 | $3.3 \times 10^{-10}$ | 24 |
| | BiFeO$_3$ crystal | 365 | 90 $\mu$m | 326 W/$m^2$ | 0.003 | $8.2 \times 10^{-12}$ | 48 |
| | BiFeO$_3$ film | 400 | 35 nm | 2 mJ/$cm^2$ | 0.46 | $4 \times 10^{-25}$ | 47 and 134 |
| | PbTiO$_3$ film | 400 | 20 nm | 4.8 mJ/$cm^2$ | 0.25 | $5.2 \times 10^{-26}$ | 52 and 134 |
| | Sn$_2$P$_2$S$_6$ crystal | Infrared | … | 210 kW/$m^2$ | $4.7 \times 10^{-7}$ | … | 35 |
| Polar semiconductors | CdS crystal | 496 | 15 $\mu$m | … | 0.0075 | … | 55 and 135 |
| | GaAs crystal | 855 | 10 $\mu$m | … | $4 \times 10^{-5}$ | … | 56 |
| Semiconductors | Germanium crystal | 1069 | 0.5 mm | … | $7.84 \times 10^{-8}$ | … | 59 |
| | Silicon crystal | 248 | 0.5 mm | 127 mJ/$cm^2$ | $-6.4 \times 10^{-4}$ | $-3.7 \times 10^{-20}$ | 66 and 136 |
| | Carbon nanotubes (MWNT) | 675 | ∼10 nm | 15 kW/$m^2$ | $1.1 \times 10^{-4}$ | $7.3 \times 10^{-19}$ | 71 |
| Chalcogenide glasses | As$_{40}$Se$_{25}$S$_{25}$Ge$_{10}$ film | … | 2.6 $\mu$m | … | 0.045 | … | 74 |
| | As$_2$Se$_3$ film | … | … | 400 W/$m^2$ | 6.4 | … | 75 |
| | As$_2$S$_3$ film | … | … | 400 W/$m^2$ | 5.4 | … | 75 |
| | As$_{40}$Se$_{50}$Ge$_{10}$ film | … | 1.6 $\mu$m | … | 0.19 | … | 74 |
| | GeSe$_2$ film | … | … | 400 W/$m^2$ | −5.6 | … | 75 |
| | GeS$_2$ film | … | … | 400 W/$m^2$ | −11 | … | 75 |
| Organic polymers | Poly-(4,4′-diamino-azobenzen-epyromellitimide) films | 400 | 0.1 mm | … | −1.2 | … | 94 |
| | Nematic elastomers | 365 | … | … | 20 | … | 96 |
| | Poly(ethylacrylate) networks with azo-aromatic crosslinks | 365, 436 | … | … | 0.25 | … | 95 |
| SC molecular crystals | [Fe(phen)$_2$(NCS)2] crystal | … | … | … | 1.1 (Ref. 137) | … | 104 |

Due to the better insulating nature and domain structure, the ferroelectric materials, however, can also be ferroelastic, meaning that a remnant strain levels induced by light can be created, involving even irreversible domain dynamics, at subcoercive electric fields. Therefore, a photoferroelastic effect can be considered as a third possible contribution to photostriction in photovoltaic ferroelectrics, involving an even larger deformation effect than expected by theory.

The photostriction in nonpolar semiconductors is generally determined by the amount of light generated charges.[138] The effect is relatively smaller and depends on the sign of the energy gap pressure susceptibility. The positive or negative susceptibilities mean positive or negative photostriction, respectively. In amorphous semiconductors, however, the excess of light-generated charges can induce plastic deformation on the atomic bonds leading to much larger photostriction. Although chalcogenide glasses show large photostriction, its plastic nature and long response time are quite limiting factors for most technical implementations besides hologram recording. The reported electron-phonon-mediated photostriction in the multiwall carbon nanotubes fits well into the class of nonpolar photostrictive semiconductors, with an efficiency better than the one of silicon. In combination with thermal effects in composite structures, the carbon nanotubes manifest an important photoactuation application potential.[139]

The photostriction in organic materials accounts for many prominent examples with large effects, where a core mechanism is either light-induced chemical structure transitions or different kinds or molecular reorientation (cooperative motion and alignment) under light irradiation.

While the largest photostriction is observed in organic based materials, their response time is generally slow limiting technical applications. However, photovoltaic ferroelectrics have an opposite problem: the response time is fast but the magnitude of the effect is small. In that respect, uniting the both advantages in a single material seems to be a promising research undertaking. Hypothetically, one can look for materials that are ferroelectric-photovoltaic and organic-based at the same time. To address this issue, one can focus on metal-organic systems, some of them were indeed reported to be ferroelectric.[140,141] Another promising candidate for photostriction can be molecular crystals; some of them were reported to be highly light-sensitive and electrically polar at the same time.[142] The spin crossover molecular systems are particularly interesting, because photostriction is accompanied with a change in magnetic properties. Thus, it might be interesting to study magnetostrictive-photostrictive coupling in some recently reported compounds.[143]

As far as ferroelectric materials are concerned, it appears to be extremely important to distinguish between the photostriction in saturated and spontaneous ferroelectric states. Based on recent experiments, the analogy to remnant strain effects[53,54] can also be seen in photostriction.[128] From a practical and short-term perspective, it appears to be the most promising to test creating ferroelastic remnant states by light and erasing them by electric fields. This can open an avenue for the light-controlled logic and memory applications in both optics and magnetic straintronics with photostrictive elements.